\documentstyle[12pt]{article}
\def\zid{1\kern-0.36em\llap~1}

\newcommand{\beq}{\begin{equation}}
\newcommand{\ber}{\begin{eqnarray}}
\newcommand{\eeq}{\end{equation}}
\newcommand{\eer}{\end{eqnarray}}

\begin{document}

\begin{titlepage}
\rightline{[SUNY BING 10/1/6 v.4 ] } \rightline{
math-ph/0610075}\vspace{2mm}
\begin{center}
{\bf \hspace{0.1 cm} Hanbury-Brown and Twiss Intensity Correlations  \\
  of Parabosons } \\
\vspace{2mm} Charles A. Nelson\footnote{Electronic address:
cnelson @ binghamton.edu  } and Paresh R. Shimpi \\ {\it
Department
of Physics, State University of New York at Binghamton\\
Binghamton, N.Y.
13902}\\[2mm]
\end{center}


\begin{abstract}
This paper shows that in intensity correlation measurements there
will be clear and unambiguous signals that new-physics particles
are, or aren't, parabosons.  For a parabosonic field in a dominant
single-mode, there is a diagonal P$_{p}$-representation in the
$|\alpha _{even,odd}>$ coherent state basis.  It is used to
analyze zero-time-interval intensity correlations of parabosons in
a maximum-entropic state. As the mean number of parabosons
decreases, there is a monotonic reduction to $\frac{2}{p}$ of the
constant bosonic ``factor of two'' proportionality of the
second-order versus the squared first-order intensity correlation
function.
\end{abstract}

\end{titlepage}

Statistics based correlations, in particular second and
higher-order intensity correlations [1,2], have proven to be very
useful in interferometry measurements and in dynamical
investigations of new phenomena in areas ranging from astrophysics
to nuclear and high energy physics [3]. It has been shown that
relativistic, local quantum field theory allows particles which
obey parastatistics, an elegant generalization of bose and fermi
statistics based on the permutation group [4]. While a use for
parastatistics in the description of natural phenomena has yet to
be found, a wealth of new but poorly studied discoveries have been
recently made in astrophysics/cosmology including an inflation
era, dark matter, and dark energy.  New particles are expected to
be produced in experiments at the Large Hadron Collider such as
Higgs particles associated with electroweak symmetry breaking
(EWSB) and supersymmetry particles associated with the
energy-scale inequality $\Lambda_{EWSB} << \Lambda_{Planck}$.
Therefore, for incisive analyses, it is both timely and important
to know both the selection rules and the novel statistical
effects/correlations [4,5] for the purpose of detecting the
presence of effective or underlying fundamental quanta obeying
parastatistics.  This paper shows that by intensity correlation
measurements there will be definitive signals that new-physics
particles are, or aren't, parabosons.

For single-mode parabosons (pB's), we introduce an index
$i=even,odd$ . The single-mode pB commutation relations [4,6] are
$ \lbrack a,\{a^{\dagger },a\}]=2a  $ and its adjoint, with
$\widehat{N}_{B}\equiv \widehat{H}_{B}-\frac{p}{2}\equiv \frac{1}{2}\{a^{\dagger },a\}-\frac{p}{2%
}$. In the occupation number basis, the states are
\begin{equation}
|n_{i} > = \frac{1}{\sqrt{(n_{i})_{p}!}}(a^{\dagger })^{n_{i}}|0>
\end{equation}
with $ n_{e} = 2N,n_{o}=2N+1;N=0,1,\ldots $; and where the
$p$-factorials are defined
\begin{equation}
\begin{array}{l}
(n_{e})_{p}! = (2N)_{p}!\equiv 2N(2N-2+p)(2N-2)\cdots 4(2+p)2p \,
;\,\,(0)_{p}!\equiv
1 \\
\hskip 3pc
 = (2N)!!(2N-2+p)!! \\
\hskip 3pc  = \{2^{N}N!\}\{2^{N}\frac{\Gamma
(N+\frac{p}{2})}{\Gamma (\frac{p}{2})}\}
\end{array}
\end{equation}
\begin{equation}
\begin{array}{l}
(n_{o})_{p}! = (2N+1)_{p}!\equiv(2N+p)2N(2N-2+p)\cdots 4(2+p)2p \\
\hskip 3pc = (2N)!!(2N+p)!! \\
\hskip 3pc  = \{2^{N}N!\}\{2^{N+1}\frac{\Gamma (N+1+\frac{p}{2})}{\Gamma (\frac{p}{2})}%
\}
\end{array}
\end{equation}
\newline
Note $(0)_{p}! = 1$, $(1)_{p}! = p$, and $(n_{o})_{p}! =
(2N+1)_{p}! = (2N+p) (2N)_{p}!$. Mathematically, in counting the
$n_p$ integers, to reach the next odd (even) $n_p$ integer, one
adds $p (1)$ to the usual integer. For fixed-order $p=1,2,\ldots $
parabosons, this $(n)_{p}!$ product of $n$ factors is analogous to
the Gibbs-factorial $n!$ for ordinary bosons ($p=1$). For creation
of one more pB, the next to vacuum position has $p (1)$ openings
when n$_{even}$ (n$_{odd}$) parabosons are already present; the
other positions
have 1 opening.\footnote{ This characterization follows from   $%
[a,(a^{\dagger })^{2n}]=2n(a^{\dagger })^{2n-1}$ and
$[a,(a^{\dagger })^{2n+1}]=(a^{\dagger })^{2n}(2n+[a,a^{\dagger
}])$ since $aa^{\dagger }|0>=p|0>$\ .} \ To make an odd number of
$a^{\dagger }$ insertions depends on $p$ but to make an even
number does not. \ Because $a^{\dagger },a$ are raising and
lowering operators, all pB single-mode $a^{\dagger },a$
commutation relations in this $ |n_{e,o} > $ basis follow from
(1).

For single-mode parafermions (pF's), in the occupation number
basis, there is a band of \ ``$p+1$'' \ states.  The pF
commutation relations are $ \lbrack c,[c^{\dagger },c]]=2c $ and
its adjoint, with
$\widehat{N}_{F}=\widehat{H}_{F}+\frac{p}{2}\equiv
\frac{1}{2}[c^{\dagger },c]+\frac{p}{2}$.  The pF number states
are equivalently
\begin{equation}
|n>_{F}=\frac{1}{\sqrt{\{n\}_{p}!}}(c^{\dagger })^{n}|0>_{F}=\frac{1}{\sqrt{%
\{p-n\}_{p}!}}(c)^{p-n}|p>_{F}; \, 0\leq n \leq p
\end{equation}
with $c|0>_{F}=c^{\dagger }|p>_{F}=0$. Analogous to the $(n)_{p}!$
for pB's, for pF's
\begin{equation}
\begin{array}{l}
\{n\}_{p}! \equiv \{n(p-n+1)\}\{(n-1)(p-n+2)\}\cdots \{2(p-1)\}\{1\,p\} \\
\hskip 3pc = n!\frac{p!}{(p-n)!}
\end{array}
\end{equation}
which is a product of $n$ bi-factors. \ Note $\{0\}_{p}!=1$ and
$\{1\}_{p}!=p$. \ In constructing a specific $|n>_{F}$ state from
the adjacent $|n-1>_{F}$ state, the additional $\{n(p-n+1)\}$
bi-factor is the product of the number of ways to insert the extra
`` $c^\dagger$ " in the first expression in (4), times the number
of ways to remove a `` $c$ " in the second expression. The
Hamiltonian defines the direction $|0>_{F}$ to $|p>_{F}$.

There are important consequences of $(n)_{p}!$ and $\{n\}_{p}!$:
For a number ``$n_{I}$'' of parabosons initially in the mode, we
let $Prob_B(\frac{a}{n_{I}}\rightarrow \frac{b}{n_{F}})$ be the
probability that in some fixed time interval a system is found to
make a transition\footnote{ A``soft accent" denotes a single
para-particle.} from a state ``$a$'' to a state ``$b$'' by the
emission or absorption of a single pB $\check{\gamma}$ . \ For
initially $n_{even}=2N$ pB's, the decay probability is \
\begin{equation}
Prob_B(\frac{a}{2N}\rightarrow \frac{b}{2N+1})=(2N+p)\ A
\end{equation}
whereas for excitation from the state ``$b$'' to ``$a$'', in the
same time interval, the absorption probability is
\begin{equation}
Prob_B(\frac{b}{2N}\rightarrow \frac{a}{2N-1})=(2N)\ A
\end{equation}
By time-reversal invariance, the constant $A$ is the same for
emission and
absorption. Hence, versus ordinary bosons, for the system initially in an $%
n_{even}$-mode, there is a $p$-dependent enhanced $\check{\gamma}$
stimulated-emission. Likewise, for initially $n_{odd}=2N+1$ pB's,
for emission/absorption \ \
\begin{equation}
\ Prob_B(\frac{a}{2N+1}%
\rightarrow \frac{b}{2N+2})=(2N+2)\ A; \, \, \,
Prob_B(\frac{b}{2N+1}\rightarrow \frac{a}{2N})=(2N+p)\ A
\end{equation}
Hence, for ordinary bosons, $p=1$, there is stimulated emission.
However, for the system initially in an  $n_{odd}$-mode, for $p=2$
the probabilities are equal; \ but for $p>2$ pB's there is a
$p$-dependent enhanced $\check{\gamma} $ absorption versus
emission. \ In summary, from consideration of single-mode pB
statistics versus bose statistics, for a system in an initial pB $n_{even}$-mode ( $%
n_{odd}$-mode ), there is a $p$-dependent enhanced
$\check{\gamma}$ stimulated-emission by the system
(stimulated-absorption by the system); otherwise there is no
$p$-dependence in these transition probabilities.  This is
different from the generic bunching signatures of ordinary bosons!

Also, in contrast, for parafermions the statistics-favored
transition probabilities are towards the ``half-full/half-empty"
$|n_{mid}>$ parafermion-state(s) at the middle of the pF band of
``$p+1$" states, due to the bi-factors and time-reversal
invariance.  From $\{n\}_{p}!$, for $p_{even}$ there is a single $%
|n_{mid}>=|\frac{p_{even}}{2}>$ state but for $p_{odd}$ there is a
pair of mid-band states $|n_{mid,high/low}=\frac{p_{odd}\pm
1}{2}>$.  For ordinary fermions, there are only these two mid-band
states:

For a number ``$n_{I}$'' of pF's initially in the mode, we let $%
Prob_{F}(\frac{a}{n_{I}}\rightarrow \frac{b}{n_{F}})$ be the
probability that in some fixed time interval a system is found to
make a transition from a state ``$a $'' to a state ``$b$'' by the
emission or absorption of a single pF $\check{\nu}$ . \ If
initially  $n_{I}>0$ is in the less-than-half-filled part of the pF band with $%
n_{I}<(\frac{p_{even}}{2},\frac{p_{odd}-1}{2})$, then the ratio of
probabilities for emission of a single  pF $\check{\nu}$ by the
system versus $\check{\nu}$ absorption is greater than one,   \
\begin{equation}
\frac{Prob_{F}(\frac{a}{n_{I}}\rightarrow \frac{b}{n_{I}+1})}{Prob_{F}(\frac{%
b}{n_{I}}\rightarrow \frac{a}{n_{I}-1})}=\frac{(n_{I}+1)(p-n_{I})}{%
n_{I}(p-n_{I}+1)}>1;\
0<n_{I}<(\frac{p_{even}}{2},\frac{p_{odd}-1}{2})\
\end{equation}
On the other hand, if initially $n_{I}<p$ is in the
more-than-half-filled part of the pF band with
$n_{I}>(\frac{p_{even}}{2},\frac{p_{odd}+1}{2})$, then the ratio
of probabilities for emission of a single pF $\check{\nu}$ by
the system versus $%
\check{\nu}$ absorption is less than one,
\begin{equation}
\frac{Prob_{F}(\frac{a}{n_{I}}\rightarrow \frac{b}{n_{I}+1})}{Prob_{F}(\frac{%
b}{n_{I}}\rightarrow \frac{a}{n_{I}-1})}=\frac{(n_{I}+1)(p-n_{I})}{%
n_{I}(p-n_{I}+1)}<1;\
p>n_{I}>(\frac{p_{even}}{2},\frac{p_{odd}+1}{2})\
\end{equation}
As the pF-mode approaches the the middle of the pF band, this is
monotonically  a smaller statistics effect. Simultaneously, if one
ignores the differences in the $A$ factors, the $%
\check{\nu}$ transition rates themselves increase as the pF-mode
approaches mid-band since the mid-band versus end-of-band ratios
\begin{equation}
\begin{array}{l}
\frac{Prob_{F}(\frac{a}{n_{mid}}\rightarrow \frac{b}{n_{mid}\pm 1})}{%
Prob_{F}(\frac{a}{n_{end}}\rightarrow
\frac{b}{n_{neighbor}})}{\biggr|}_{statistics\
factor\ only} = \frac{1}{4}(p_{even}+2)>1;\ p_{even}>2 \\
\\
\frac{Prob_{F}(\frac{a}{n_{mid,low}}\leftrightarrow \frac{b}{n_{mid,high}})}{%
Prob_{F}(\frac{a}{n_{end}}\rightarrow
\frac{b}{n_{neighbor}})}{\biggr|}_{statistics\ factor\ only} =
\frac{1}{4}(p_{odd}+2+\frac{1}{p_{odd}})>1;\ p_{odd}>1
\end{array}
\end{equation}

The pB coherent states [6] can be written in terms of a
$p$-exponential function
\begin{equation}
\begin{array}{l}
e_p(x) \equiv \sum\limits_{n=0}^{\infty }\frac{x^{n}}{(n)_{p}!} \\
\hskip 3pc = e_{e}(x)+e_{o}(x);\,\,e_{e,o}(-x)=\pm e_{e,o}(x).
\end{array}
\end{equation}
In terms of the modified Bessel function $I_{\nu }(x)$, the even
and odd terms are
\begin{equation}
e_{e}(x)=(\frac{x}{2})^{\frac{2-p}{2}}\Gamma (\frac{p}{2})I_{\frac{p-2}{2}%
}(x);e_{o}(x)=(\frac{x}{2})^{\frac{2-p}{2}}\Gamma (\frac{p}{2})I_{\frac{p}{2}%
}(x)\,\,
\end{equation}
For $p=1$, $e_{e,o}(x) \rightarrow \cosh x , \sinh x $.

Thereby, with $\alpha $ complex-valued, $a|\alpha
>=\alpha |\alpha
>$ for
\begin{equation}
\begin{array}{l}
|\alpha  > = \frac{1}{\sqrt{e_p(|\alpha |^{2})}}e_p(\alpha \, a^{\dagger })|0> \\
\hskip 2pc  =\sqrt{P_{e}(|\alpha |^{2})}\,|\alpha
_{e}>+\sqrt{P_{o}(|\alpha |^{2})}\,|\alpha _{o}>
\end{array}
\end{equation}
with
\begin{equation}
\begin{array}{l}
\,|\alpha _{e} > =\frac{1}{\sqrt{e_{e}(|\alpha |^{2})}}\sum\limits_{N=0}^{\infty }%
\frac{|\alpha |^{2N}}{(2N)_{p}!}|2N> \\
\,|\alpha _{o} > =\frac{1}{\sqrt{e_{o}(|\alpha |^{2})}}\sum\limits_{N=0}^{\infty }%
\frac{|\alpha |^{2N+1}}{(2N+1)_{p}!}|2N+1> \\
<\alpha _{e}|\alpha _{e}>=<\alpha _{o}|\alpha _{o}>=1;<\alpha
_{e}|\alpha _{o}>=0
\end{array}
\end{equation}
In (14), the important $n_{e,o}$ coherent-state-mode probabilities
are
\begin{equation}
P_{e,o}(|\alpha |^{2})\equiv \frac{e_{e,o}(|\alpha
|^{2})}{e_p(|\alpha |^{2})}
\end{equation}
These range monotonically in $ | \alpha |^2 $ from $P_{e,o}
\rightarrow 1- \frac{| \alpha |^2}{p}, \frac{| \alpha |^2}{p}$
respectively as $ | \alpha |^2 \rightarrow 0$, to $P_{e,o}
\rightarrow \frac{1}{2}$ for $ | \alpha |^2
>> 1, \frac{p}{2} $.

Although $a^{2}|\alpha _{e,o}>=\alpha ^{2}|\alpha _{e,o}>$, it is
with a different normalization
\begin{equation}
|\alpha _{+}>\equiv \sqrt{2P_{e}(|\alpha |^{2})}\,|\alpha
_{e}>;\,\,|\alpha _{-}>\equiv \sqrt{2P_{o}(|\alpha
|^{2})}\,|\alpha _{o}>
\end{equation}
that $a|\alpha _{\pm }>=\alpha |\alpha _{\mp }>$. \ Note that
$|\alpha _{\pm }>=\frac{1}{\sqrt{2}}[|\alpha >\pm \,|-\alpha >]$,
\newline $|-\alpha >=\sqrt{%
P_{e}(|\alpha |^{2})}\,|\alpha _{e}>-\sqrt{P_{o}(|\alpha
|^{2})}\,|\alpha _{o}>$, and $ |\alpha
_{e,o}>=\frac{1}{\sqrt{2P_{e,o}(|\alpha |^{2})}}[|\alpha
>\pm \,|-\alpha >] $.

There is an associated $p$-Poisson distribution function, $ x=|
\alpha |^{2} $,
\begin{equation}
{\cal P}_{p}(n_{i},x)\equiv
\frac{x^{n_{i}}}{(n_{i})_{p}!}\frac{1}{e_p(x)} = |<n_{i}|\,\alpha
>|^{2}; \, \,  i= even,odd
\end{equation}
for the probability to be in the $n$th number-state if the system
is in the coherent state $|\alpha >$.
\begin{equation}
\begin{array}{c}
\,  {\cal P}_{p} (n_{i},x) \rightarrow \frac{|\alpha
|^{2n_{i}}}{(n_{i})_{p}!}\{1-\frac{|\alpha
|^{2}}{p}+{\cal{O}}(|\alpha |^{4})\},\ \ |\alpha |^{2}\rightarrow 0 \\
\hskip 14pc \rightarrow \frac{2^{\frac{p-1}{2}}\sqrt{\pi }}{\Gamma (\frac{p}{2})}[%
\frac{|\alpha |^{2(n_{i}-[\frac{p-1}{2}])}}{(n_{i})_{p}!}\exp
(-|\alpha |^{2})]\{1+{\cal{O}}(\frac{1}{|\alpha |^{2}})\},\ \
|\alpha |^{2}>>1,\frac{p}{2}
\end{array}
\end{equation}
The corresponding $p$-Gaussian ( $p$-normal ) distribution
approximation to (18) is given in the appendix.  The
$p$-dependence in the $p$-Gaussian distribution for the pB
coherent state arises only through the mean $\mu$ and the standard
deviation $\sigma $. This is unlike the explicit $p$-dependence,
and the explicit $n_{even}$ versus $n_{odd}$ differences, in the
above $p$-Poisson distribution. From (18),$\frac{{\cal
P}_{p}(2N+1)}{{\cal P}_{p}(2N)}=\frac{|\alpha |^{2}}{2N+p}$ and
$\frac{{\cal P}_{p}(2N-1)}{{\cal P}_{p}(2N)}=\frac{2N}{|\alpha
|^{2}}$.  This means for the distribution of $| n
> $'s in the coherent state $|\alpha>$ that versus the probability of
an arbitrary  $| n_{even} > $ number-state, the next ``odd"
number-state is less probable for $p>1$ than for $p=1$, but the
probability for the previous ``odd" is not $p$-dependent.

For two arbitrary coherent states,
\begin{equation}
<\alpha |\beta >=\sqrt{P_{e}(|\alpha |^{2})P_{e}(|\beta
|^{2})}<\alpha _{e}|\beta _{e}>+\,\,\sqrt{P_{o}(|\alpha
|^{2})P_{o}(|\beta |^{2})}<\alpha _{o}|\beta _{o}>
\end{equation}
with
\begin{equation}
<\alpha _{e}|\beta _{e}>=\frac{e_{e}(\alpha ^{\ast }\beta )}{\sqrt{%
e_{e}(|\alpha |^{2})e_{e}(|\beta  |^{2})}};\,\,<\alpha _{o}|\beta
_{o}>=\frac{e_{o}(\alpha ^{\ast }\beta )}{\sqrt{e_{o}(|\alpha
|^{2})e_{o}(|\beta |^{2})}};\,\,<\alpha _{e}|\beta _{o}>=0
\end{equation}
So for large arguments: $|\alpha |,|\beta |>>1,\frac{p-2}{2},$%
\begin{equation}
|<\alpha |\beta >|\rightarrow \exp (-\frac{1}{2}|\alpha -\beta |^{2})\{1+{\cal{O}}(\frac{1}{%
|\alpha |^{2}},\frac{1}{|\beta |^{2}},\frac{1}{|\alpha ^{\ast
}\beta |})\}
\end{equation}
and for small arguments: $|\alpha |,|\beta |<<1,$%
\begin{equation}
|<\alpha |\beta >|\rightarrow \{1-\frac{|\alpha -\beta
|^{2}}{2p}+{\cal{O}}(|\alpha |^{4},|\beta |^{4},|\alpha
|^{2}|\beta |^{2})\}
\end{equation}
As a consequence of the completeness relation
\begin{equation}
I\equiv \frac{1}{\pi }\int d^{2}\alpha \,\,\mu _{e}(|\alpha
|^{2})\,|\alpha _{e}><\,\alpha _{e}|\,+\frac{1}{\pi }\int
d^{2}\alpha \,\,\mu _{o}(|\alpha |^{2})\,|\alpha _{o}><\,\alpha
_{o}|\, ,
\end{equation}
where in terms both types of modified Bessel functions
\begin{equation}
\mu _{e}(|\alpha |^{2}) = |\alpha |^{2}K_{\frac{p-2}{2}}(|\alpha |^{2})I_{%
\frac{p-2}{2}}(|\alpha |^{2});\,\,\mu _{o}(|\alpha |^{2})=|\alpha |^{2}K_{%
\frac{p}{2}}(|\alpha |^{2})I_{\frac{p}{2}}(|\alpha |^{2}),
\end{equation}
and of the $|\alpha >$ and $|\beta >$ non-orthogonality, the pB
coherent states are linearly dependent.  They are overcomplete.
This relation\footnote{ By substitution of (17) into (24),
eq(2.63) of 3rd paper in [6] is obtained. } (24) follows by using
\begin{equation}
\begin{array}{l}
{\Gamma_e}( n_e +1 ) \equiv (2N)_{p}! =
\frac{2^{\frac{2-p}{2}}}{\Gamma (\frac{p}{2})}\int_{0}^{\infty
}d(|\alpha |^{2})K_{\frac{p-2}{2}}(|\alpha |^{2})|\alpha |^{2(\frac{p}{2}%
+n_e)} \\
{\Gamma_o}( n_o +1 ) \equiv (2N+1)_{p}! = \frac{2^{\frac{2-p}{2}}}{\Gamma (\frac{p}{2})}%
\int_{0}^{\infty }d(|\alpha |^{2})K_{\frac{p}{2}}(|\alpha |^{2})|\alpha |^{2(%
\frac{p}{2}+ n_o )}
\end{array}
\end{equation}
This analytic integral representation for the two $p$-factorials
generalizes Euler's formula for $\Gamma (x)$.  Uses for
generalizations of other functions of integers, e.g.
$\zeta_{p}(k)\equiv \sum\limits_{n_{p}=p}^{\infty }(n_{p})^{-k}$,
$k>1$ , remain to found.

In this $|\alpha _{e,o}>$ coherent state basis, there is a
diagonal P$_{p}$-representation\footnote{This is the direct
generalization of the bosonic P-representation, see [7]. We use
$\Phi^{(e,o)}(\alpha ) $ to denote the weight functions, so as to
avoid confusion with other functions.} for the density operator
$\widehat{\rho }$ describing the state of the system
\begin{equation}
\widehat{\rho } \equiv \frac{1}{\pi }\int d^{2}\alpha \,\,\mu
_{e}(|\alpha |^{2})\,|\alpha _{e}><\,\alpha
_{e}|\,\Phi^{(e)}(\alpha )+\frac{1}{\pi }\int d^{2}\alpha \,\,\mu
_{o}(|\alpha |^{2})\,|\alpha _{o}><\,\alpha
_{o}|\,\Phi^{(o)}(\alpha )
\end{equation}
For $\widehat{\rho }=\widehat{\rho }^{\dagger }$,
$\Phi^{(e,o)}(\alpha )$ are
real. \ The normalization condition from $\widehat{\rho }=\widehat{\rho }_{e}+%
\widehat{\rho }_{o}$ is
\begin{equation}
{ Tr}\,\widehat{\rho }_{e,o}=I_{e,o}=\frac{1}{\pi }\int
d^{2}\alpha \,\,\mu _{e,o}(|\alpha |^{2})\,\Phi^{(e,o)}(\alpha )
\end{equation}
For instance, if the system is in the coherent state $|\beta_{e}>$ , $\ $%
then $\widehat{\rho}_{\beta _{e}}=|\beta _{e}><\,\beta _{e}|$ for\ $\Phi^{(e)}(\alpha )=%
\frac{\pi }{\mu _{e}(|\beta |^{2})} \delta^{2}( \alpha - \beta
),\Phi^{(o)}(\alpha )=0$ where $ \delta ^{2}( \alpha - \beta ) =
\delta ( {Re} [  \alpha -\beta ])\delta ( {Im}[\alpha -\beta ])$.

To describe a field theoretic system in a maximum-entropic state,
we proceed as in the treatment of ordinary bosons [2] for which
such a field-state is often called a chaotic or thermal state: In
terms of the paraboson number operator
$\widehat{N}=\widehat{N}_{B}$, we maximize the entropy $S=-k \,
Tr(\rho  \ln \rho )$ to obtain
\begin{equation}
\widehat{\rho }_{\max S}=\frac{1}{1+\,<\widehat{N}>}\left( \frac{<\widehat{N}%
>}{1+\,<\widehat{N}>}\right) ^{\widehat{N}}
\end{equation}
where $<\widehat{N}>$ is the mean number of parabosons in the
maximum-entropic state.

Defining $r \equiv 1+\frac{1}{<\widehat{N}>}$, from the above
P$_{p}$-representation, the corresponding non-negative maximum-entropic $%
\Phi^{(e,o)}(\alpha )$ functions are
\begin{equation}
\Phi _{\max S}^{(e)}(\alpha )=\frac{r^{\frac{p}{2}}}{<\widehat{N}>}%
\frac{K_{\frac{p-2}{2}}(r |\alpha
|^{2})}{K_{\frac{p-2}{2}}(|\alpha
|^{2})};\,\,\Phi _{\max S}^{(o)}(\alpha )=\frac{r^{\frac{p}{2}}}{<%
\widehat{N}>}\frac{K_{\frac{p}{2}}(r |\alpha |^{2})}{K_{\frac{p}{2}%
}(|\alpha |^{2})}
\end{equation}
For ordinary bosons $\Phi _{\max S}^{(e)}(\alpha )=\Phi _{\max
S}^{(o)}(\alpha )=\frac{1}{<\widehat{N}>}\exp (-|\alpha |^{2}/<\widehat{N}>)$%
.

By using these results, in analogy with the scalar-field treatment
for ordinary bosons, we can analyze Hanbury-Brown and Twiss
intensity correlations [1,2] for a paraboson field which is in a
single-mode maximum-entropic state: \ We consider
zero-time-interval correlations. \ Using $\widehat{\rho }_{\max
S}=\widehat{\rho }_{\max S}^{(e)}+\widehat{\rho }_{\max S}^{(o)}$
, the first-order coherence function is
\begin{equation}
\begin{array}{l}
G^{(1)}(0) \equiv Tr[\widehat{\rho }_{\max
S}\,\,E^{(-)} ( x_{1} ) E^{(+)}(x_{1})] \\
\hskip 3pc = G_{e}^{(1)}(0)+G_{o}^{(1)}(0) \\
\hskip 3pc = \overline{c}<\widehat{N}>\left[ \frac{p+2<\widehat{N}>}{1+2<\widehat{N}>}%
\right]
\end{array}
\end{equation}
where $\overline{c}$ is a constant factor. \ The second-order, or
intensity, correlation function is\begin{equation}
\begin{array}{l}
G^{(2)}(0) \equiv Tr[\widehat{\rho }_{\max
S}\,\,E^{(-)}(x_{1})E^{(-)}(x_{2})E^{(+)}(x_{2})E^{(+)}(x_{1})] \\
\hskip 3pc = 2\,\,(\overline{c}<\widehat{N}>)^{2}\left[ \frac{p+2<\widehat{N}>}{1+2<%
\widehat{N}>}\right]
\end{array}
\end{equation}
If we write a proportionality $G^{(2)}(0)=\ \lambda
_{p}\,[G^{(1)}(0)]^{2}$,
then $\lambda _{p}=2\,\,\left[ \frac{1+2<\widehat{N}>}{p+2<\widehat{N}>}%
\right] $ which shows that as the mean number of parabosons
decreases, there is a monotonic reduction to $\frac{2}{p}$ versus
the usual bosonic ``constant factor of two'' intensity correlation
effect.

Similarly, we obtain for the next higher-order correlation
functions
\begin{eqnarray}
G^{(3)}(0)= 3!(\overline{c}<\widehat{N}>)^{3}\left[ \frac{a\,p^{2}+b\,p+c}{%
3(1+2<\widehat{N}>)^{3}}\right] \\
G^{(4)}(0) = 4!(\overline{c}<\widehat{N}>)^{4}\left[ \frac{a\,p^{2}+b\,p+c}{%
3(1+2<\widehat{N}>)^{3}}\right]
\end{eqnarray}
where\begin{equation}
\begin{array}{l}
a = 1+2<\widehat{N}> \\
b = 2(1+4<\widehat{N}>+6<\widehat{N}>^{2}) \\
c = 8<\widehat{N}>(1+3<\widehat{N}>+3<\widehat{N}>^{2})
\end{array}
\end{equation}
As in (31,32), the last factor in (33,34) depends on $<N>$; it
approaches ``1'' for $<N>$ large, \ and a p-dependent value ``
$\frac{p^{2}+2p}{3} $ '' as $ <N> \rightarrow 0 $.

Formulae for arbitrary $n$ order ${ { G_{e,o} }^{(n)} } (0)$ are
listed in the appendix. There is no $p$-dependence in
zero-time-interval, intensity correlation functions in the large
$<\widehat{N}>$ limit. However, there is significant
$p$-dependence as the mean number of parabosons decreases.

In summary, this paper shows that in intensity correlation
measurements there will be clear and unambiguous signals that new
quanta are, or aren't, parabosons.  It will be a complete
measurement, because the order $p$ of their parastatistics will be
determined!

One of us (CAN) thanks colleagues at Binghamton University and
elsewhere, in physics and in mathematics, for discussions. This
work was partially supported by U.S. Dept. of Energy Contract No.
DE-FG 02-86ER40291.

\bigskip

\bigskip

{\bf Note added in proof:  Recursion relation for intensity
correlation functions:}

There is a simple recursion relation between the $n_{even}$-order
and the lower-adjacent $n_{odd}$-order intensity correlation
function
$$ G^{(n_e)}(0)= \ n_e \
(\overline{c}<\widehat{N}>) \ G^{(n_e -1)}(0).
$$
This relation is independent of $p$ and so it might be empirically
very useful, for instance for measurement of the product of the
constant factor and the mean number of parabosons, $
(\overline{c}<\widehat{N}>) $. This relation follows from the
hypergeometric function expressions in the appendix.

\newpage

{\bf Appendix:}

{\bf  $ G^{(n)} (0) $ Formulae for Maximum-Entropic State:}

For $G^{(n)}(0)=G_{e}^{(n)}(0)+G_{o}^{(n)}(0)$ with $r=1+\frac{1}{<\widehat{N%
}>}$, there are integral representations
\begin{equation}
\begin{array}{l}
G_{e}^{(n_{o})}(0)
=\frac{r^{\frac{p}{2}}}{<\widehat{N}>}\int\limits_{0}^{\infty
}dx\,\,x^{n_{o}+1}\,K_{\frac{p}{2}-1}(rx)\,I_{\frac{p}{2}}(x) \\
G_{o}^{(n_{o})}(0)
=\frac{r^{\frac{p}{2}}}{<\widehat{N}>}\int\limits_{0}^{\infty
}dx\,\,x^{n_{o}+1}\,K_{\frac{p}{2}}(rx)\,I_{\frac{p}{2}-1}(x) \\
G_{e}^{(n_{e})}(0)
=\frac{r^{\frac{p}{2}}}{<\widehat{N}>}\int\limits_{0}^{\infty
}dx\,\,x^{n_{e}+1}\,K_{\frac{p}{2}-1}(rx)\,I_{\frac{p}{2}-1}(x) \\
G_{o}^{(n_{e})}(0)
=\frac{r^{\frac{p}{2}}}{<\widehat{N}>}\int\limits_{0}^{\infty
}dx\,\,x^{n_{e}+1}\,K_{\frac{p}{2}}(rx)\,I_{\frac{p}{2}}(x)
\end{array}
\end{equation}
In this appendix, we suppress the overall $ ( \bar{c}) ^{n_i} $
factors. The $G_{e,o}^{(n_i)}(0)$ can be written in terms of the
hypergeometric function or as an infinite series
\begin{equation}
\begin{array}{l}
G_{e}^{(n_{o})}(0)
=\frac{2^{n_{o}}}{<\widehat{N}>r^{n_{o}+2}}\frac{\Gamma
(\frac{p}{2}+\frac{n_{o}}{2}+\frac{1}{2})\Gamma (\frac{n_{o}}{2}+\frac{3}{2})%
}{\Gamma (\frac{p}{2}+1)}\,\,_{2}F_{1}(\frac{p}{2}+\frac{n_{o}}{2}+%
\frac{1}{2},\frac{n_{o}}{2}+\frac{3}{2};\frac{p}{2}+1;r^{-2}) \\
\hskip 4pc =\frac{2^{n_{o}}}{<\widehat{N}>r^{n_{o}+2}}\sum\limits_{m=0}^{\infty }\frac{%
\Gamma (\frac{p}{2}+N+1+m)\Gamma (N+2+m)}{\Gamma (\frac{p}{2}+1+m)\Gamma (m+1)}%
\ r^{-2m} \\
G_{o}^{(n_{o})}(0) =
\frac{2^{n_{o}}}{<\widehat{N}>r^{n_{o}+1}}\frac{\Gamma
(\frac{p}{2}+\frac{n_{o}}{2}+\frac{1}{2})\Gamma (\frac{n_{o}}{2}+\frac{1}{2})}{%
\Gamma (\frac{p}{2})}\ _{2}F_{1}(\frac{p}{2}+\frac{n_{o}}{2}+\frac{1%
}{2},\frac{n_{o}}{2}+\frac{1}{2};\frac{p}{2};r^{-2}) \\
\hskip 4pc =\frac{2^{n_{o}}}{<\widehat{N}>r^{n_{o}+1}}%
\sum\limits_{m=0}^{\infty }\frac{\Gamma (\frac{p}{2}+N+1+m)\Gamma (N+1+m)}{\Gamma (%
\frac{p}{2}+m)\Gamma (m+1)}\ r^{-2m} \\
G_{e}^{(n_{e})}(0) =
\frac{2^{n_{e}}}{<\widehat{N}>r^{n_{e}+1}}\frac{\Gamma
(\frac{p}{2}+\frac{n_{e}}{2})\Gamma (\frac{n_{e}}{2}+1)}{\Gamma (\frac{p}{2})%
}\ _{2}F_{1}(\frac{p}{2}+\frac{n_{e}}{2},\frac{n_{e}}{2}+1;\frac{p}{2%
};r^{-2}) \\
\hskip 4pc =\frac{2^{n_{e}}}{<\widehat{N}>r^{n_{e}+1}}%
\sum\limits_{m=0}^{\infty }\frac{\Gamma (\frac{p}{2}+N+m)\Gamma (N+1+m)}{\Gamma (%
\frac{p}{2}+m)\Gamma (m+1)}\ r^{-2m} \\
G_{o}^{(n_{e})}(0) =
\frac{2^{n_{e}}}{<\widehat{N}>r^{n_{e}+2}}\frac{\Gamma
(\frac{p}{2}+\frac{n_{e}}{2}+1)\Gamma (\frac{n_{e}}{2}+1)}{\Gamma (\frac{p}{2%
}+1)}\ _{2}F_{1}(\frac{p}{2}+\frac{n_{e}}{2}+1,\frac{n_{e}}{2}+1;%
\frac{p}{2}+1;r^{-2}) \\
\hskip 4pc =\frac{2^{n_{e}}}{<\widehat{N}>r^{n_{e}+2}}%
\sum\limits_{m=0}^{\infty }\frac{\Gamma (\frac{p}{2}+N+1+m)\Gamma (N+1+m)}{\Gamma (%
\frac{p}{2}+1+m)\Gamma (m+1)}\ r^{-2m}
\end{array}
\end{equation}

For $<\widehat{N}> \rightarrow \infty $, we assume the large
$x=|\alpha |^{2}$
\ region of the integrand dominates, so $G_{e,o}^{(n)}(0)\rightarrow \frac{1%
}{2}n!\,\,<\widehat{N}>^{n}$ because of the absence of $\nu $%
-dependence in $I_{\nu }(x)$ and $ K_{\nu }(x)$ for large $x$. \

Expansions in $p$ follow from the infinite series expressions,
\begin{equation}
\begin{array}{l}
G_{e}^{(n_{o})}(0) =\frac{2^{N}N!<\widehat{N}>^{2N+1}}{(1+2<\widehat{N}%
>)^{N+2}}\{ p^{N+1}[1+2<\widehat{N}>] \cr
\hskip 12pc +p^{N}(N+1)[N+2N<\widehat{N}%
>+2(N+1)<\widehat{N}>^{2}]+\cdots \}  \\
G_{o}^{(n_{o})}(0) =\frac{2^{N}N!<\widehat{N}>^{2N+1}}{(1+2<\widehat{N}%
>)^{N+2}}\{ p^{N}(N+1)[2<\widehat{N}>+2<\widehat{N}>^{2}]+\cdots
\}  \\
G^{(n_{o})}(0) =\frac{2^{N}N!<\widehat{N}>^{2N+1}}{(1+2<\widehat{N}>)^{N+2}%
}\{ p^{N+1}[1+2<\widehat{N}>] \cr
\hskip 12pc +p^{N}(N+1)[N+2(N+1)<\widehat{N}>+2(N+2)<%
\widehat{N}>^{2}]+\cdots \}
\end{array}
\end{equation}
and
\begin{equation}
\begin{array}{l}
G_{e}^{(n_{e})}(0) =\frac{2^{N}N!<\widehat{N}>^{2N+1}}{(1+2<\widehat{N}%
>)^{N+2}}\{ p^{N}[1+2<\widehat{N}>] \cr
\hskip 12pc +p^{N-1}(N)[N+1+2(N+1)<\widehat{N}%
>+2(N+1)<\widehat{N}>^{2}]+\cdots \}  \\
G_{o}^{(n_{e})}(0) =\frac{2^{N}N!<\widehat{N}>^{2N}(1+<\widehat{N}>)}{(1+2<%
\widehat{N}>)^{N+2}}\{ p^{N}[1+2<\widehat{N}>] \cr
\hskip 12pc +p^{N-1}(N)[N-1+2(N-1)<%
\widehat{N}>+2(N+1)<\widehat{N}>^{2}]+\cdots \}  \\
G^{(n_{e})}(0) =\frac{2^{N}N!<\widehat{N}>^{2N}}{(1+2<\widehat{N}>)^{N+1}}%
\{ p^{N}[1+2<\widehat{N}>] \cr
\hskip 12pc +p^{N-1}(N)[N-1+2N<\widehat{N}>+2(N+1)<%
\widehat{N}>^{2}]+\cdots \}
\end{array}
\end{equation}

{\bf $p$-Gaussian Distribution:}

Per the central limit theorem, the $p$-Gaussian distribution (
$p$-normal ) approximation to the $p$-Poisson distribution follows
from (18):

In the $p$-Poisson distribution function ${\cal P}_{p}(n_i,x)$ for
large $ x=|\alpha |^{2}$, we change variables to $y=\frac{n-\mu
}{\sigma }$, so as to measure the deviation of $n$ versus the mean
$\mu$ in units of the standard deviation $\sigma$.
\begin{equation}
\begin{array}{l}
\mu  \equiv  <\alpha |\widehat{N}|\alpha > \\
=|\alpha |^{2}+\frac{1}{2}(1-p)-\frac{D}{2}(1-p)
\end{array}
\end{equation}
\begin{equation}
\begin{array}{l}
\sigma ^{2} \equiv  <\alpha |(\widehat{N}-\mu )^{2}|\alpha > \\
=|\alpha |^{2}+\frac{1}{2}(1-p)^{2}+D(1-p)|\alpha |^{2}-\frac{1}{4}%
D^{2}(1-p)^{2}
\end{array}
\end{equation}
Both depend on the positive difference between the $n_{e,o}$
coherent-state-mode probabilities of (14)
\begin{equation}
D=D(|\alpha |^{2}) \equiv P_{e}(|\alpha
|^{2})-P_{o}(|\alpha |^{2}) = \frac {[I_{\frac{p-2}{2}}(|\alpha |^{2})-I_{\frac{p}{2}%
}(|\alpha |^{2})]} {[I_{\frac{p-2}{2}}(|\alpha
|^{2})+I_{\frac{p}{2}}(|\alpha |^{2})]} >0.
\end{equation}
For $ | \alpha |^2 >> 1, \frac{p}{2} $,
\begin{equation}
\begin{array}{l}
 \mu \simeq |\alpha
|^{2}+\frac{1}{2}(1-p) \\
\sigma ^{2}   \simeq |\alpha |^{2}+\frac{1}{2}(1-p)^{2}
\end{array}
\end{equation}
since $D\ \rightarrow 0$. For $ | \alpha |^2 \rightarrow 0$, $D
\rightarrow 1-\frac{2 | \alpha |^2}{p} + {\cal O} (| \alpha |^{4}
) $.

Expanding and using the Stirling approximation, we obtain as in
the $p=1$ case [8]
\begin{equation}
\begin{array}{l}
{\cal{P}}_{p}(n_{e},y) = \frac{1}{\sigma \sqrt{2\pi }}e^{-\frac{1}{2}%
y^{2}}[\ 1-\frac{1}{\sigma }(\frac{y}{2}-\frac{y^{3}}{6}) \cr
\hskip 12pc  -\frac{1}{\sigma ^{2}}(\frac{1}{12}+\frac{p}{4}-\frac{p^{2}}{4}-y^{2}\{%
\frac{1}{8}+\frac{p}{2}-\frac{p^{2}}{4}\}+\frac{y^{4}}{6}-\frac{y^{6}}{72})\cr
\hskip 12pc  + {\cal{O}}(\frac{1}{\sigma ^{3}})\ ]; \, \, y \equiv
\frac{n-\mu }{\sigma }
\end{array}
\end{equation}
For ${\cal{P}}_{p}(n_{o},y)$ replace `` $\frac{1}{12}+\frac{p}{4}$
'' in the $\sigma ^{-2}$ coefficient by `` $-\frac{5}{12}+\frac{3
p}{4}$ ''. In the bosonic case, additional terms through $\sigma
^{-4}$ coefficients are given in [8].

As for $p=1$, using the first term of this series, we define the
$p$-Gaussian distribution ($p$-normal)
\begin{equation}
{\cal{G}}_{p}(n,\mu ,\sigma )\equiv \frac{1}{\sigma \sqrt{ 2 \pi
}}\exp [-\frac{1}{2}(\frac{n-\mu }{\sigma })^{2}]
\end{equation}
where $\mu ,\sigma $ for the pB coherent state are given in
(40,41). Unless $\frac{p^2}{{\sigma}^2 }$ and/or $\frac{(y
p)^2}{{\sigma}^2 }$ is large, for $|\alpha |^{2}$ large the
$p$-Gaussian distribution will be a satisfactory approximation to
the $p$-Poisson when $\frac{y^3}{\sigma }$ is small.

\end{document}